# ChemGraph: An Agentic Framework for Computational Chemistry Workflows

Thang D. Pham[1], Aditya Tanikanti[2], Murat Keçeli[1]

[1]Computational Science (CPS) Division, Argonne National Laboratory, Lemont, IL

[2]Argonne Leadership Computing Facility (ALCF) Division, Argonne National Laboratory, Lemont, IL

Corresponding authors' email addresses: tpham@anl.gov, keceli@anl.gov

## Abstract

Atomistic simulations are essential tools in chemistry and materials science, accelerating the discovery of novel catalysts, energy storage materials, and pharmaceuticals. However, running these simulations remains challenging due to the wide range of computational methods, diverse software ecosystems, and the need for expert knowledge and manual effort for the setup, execution, and validation stages. In this work, we present ChemGraph, an agentic framework powered by artificial intelligence and state-of-the-art simulation tools to streamline and automate computational chemistry and materials science workflows. ChemGraph leverages graph neural network-based foundation models for accurate yet computationally efficient calculations and large language models (LLMs) for natural language understanding, task planning, and scientific reasoning to provide an intuitive and interactive interface. Users can perform tasks such as molecular structure generation, single-point energy, geometry optimization, vibrational analysis, and thermochemistry calculations with methods ranging from tight-binding and machine learning interatomic potentials to density functional theory or wave function theory-based methods. We evaluate ChemGraph across 13 benchmark tasks and demonstrate that smaller LLMs (GPT-4o-mini, Claude-3.5-haiku, Qwen2.5-14B) perform well on simple workflows, while more complex tasks benefit from using larger models like GPT-4o. Importantly, we show that decomposing complex tasks into smaller subtasks through a multi-agent framework enables smaller LLM models to match or exceed GPT-4o's performance in specific scenarios.

## Introduction

Atomistic simulations play an important role in chemistry and materials science[1–4], enabling the design of new and improved catalysts[5,6], energy storage materials[1,7,8], and accelerating drug discovery[9,10]. Techniques such as density functional theory (DFT), coupled cluster (CC) methods, molecular dynamics (MD), and Monte Carlo (MC) simulations have been widely adopted to predict molecular and material properties, explore reaction mechanisms, and optimize performance at the atomic level[11–13]. Recent advances in machine learning, particularly the development of graph neural networks (GNNs) and foundation models for molecules and materials, have enabled highly accurate and scalable alternatives to traditional quantum mechanical methods.[14–17] These models are typically trained on large datasets generated from DFT calculations, allowing them to produce DFT-level accuracy at a fraction of the computational cost. Furthermore, their fast inference times enable interactive and real-time molecular simulations, opening new possibilities for user-driven exploration in chemistry and materials research.

Running molecular simulations efficiently remains a complex and time-consuming task, often requiring specialized expertise and manual effort across the workflow. Researchers must carefully define system parameters, select appropriate computational methods, and generate input files, all tailored to their specific needs. This process is further complicated by the diversity of simulation software, each with its own input syntax, programming language, and interfaces. Even a minor mistake in the input configuration can lead to incorrect results, unnecessary computational expenses or failures. To address these challenges, Python libraries such as the Atomic Simulation Environment (ASE)[18] and QCEngine[19] provide a unified interface and act as universal executors for many simulation software. Additionally, open databases like the Materials Project[20] and Open Molecules 2025 (Omol 2025)[21], which publish both simulation inputs and outputs, promote reproducibility and help users adopt widely accepted parameters and practices across a broad range of molecular simulation tasks.

Recent advancements in artificial intelligence (AI), particularly in large language models (LLMs), have opened new possibilities for automating scientific research.[22] LLMs have demonstrated exceptional capabilities in natural language understanding, reasoning, and task execution, making them well-suited for guiding complex workflows.[23,24] Indeed, several LLM-based agents (assistants), have been developed to help users in various chemistry-related tasks.[25–29] Bran et al. developed ChemCrow, a large language model (LLM)-powered chemistry agent capable of executing chemical synthesis for a variety of molecules.[25] Additionally, ChemCrow demonstrates a human-AI collaboration framework that enhances the LLM agent's ability to discover novel molecules. McNaughton et al. developed Chemistry Agent Connecting Tool Usage to Science (CACTUS), an LLM-based agent integrated with cheminformatics tools that can assist researchers in tasks such as molecular property prediction, similarity searching, and drug-likeness assessment.[26] Recently, White and colleagues introduced MDCrow, an LLM agent assistant that can perform molecular dynamics (MD) workflow.[28] Aspuru-Guzik and colleagues released El Agente, an LLM agent for quantum chemistry calculations.[29]

In this work, we introduced ChemGraph[30], a large language model (LLM)-powered agent system designed to perform molecular simulation workflows in computational chemistry. ChemGraph integrates natural language processing with simulation tools, to perform a series of tasks ranging from SMILES string and molecular structure generation to geometry optimization, vibrational analysis and thermochemistry calculation. ChemGraph allows users to utilize a wide range of molecular simulation methods, including semi-empirical methods, ML potentials, and DFT. Our evaluation demonstrates that for simpler tasks requiring only a few tool calls, smaller LLMs such as GPT-4o-mini and Claude-3.5-haiku achieve relatively high accuracy and consistency. However, as the task complexity increases, their performance declines significantly, while a large model (GPT-4o) maintains strong performance. By strategically decomposing complex tasks into smaller, manageable subtasks, we increased the performance of ChemGraph even when using smaller models, achieving results that are comparable to, and in some cases even surpass, those of GPT-4o.

By abstracting away low-level coding and tool-specific configurations, ChemGraph allows users to perform molecular simulations using intuitive prompts. This approach not only lowers the barriers for computational chemistry research but also provides automation that can enhance high-throughput workflows.

**Methodology**

**Framework.** ChemGraph is implemented with LangGraph[31] and follows the ReAct framework[32]. LangGraph introduces a graph-based execution model designed for more robust multi-agent coordination. A typical LangGraph workflow consists of three components: state, nodes, and edges. The state is a shared data structure representing the system's status. Nodes are Python functions that define the logic of agents, taking the current state as input and returning an updated state. The nodes can be an LLM or a function executing a specific operation. Edges are functions that control the flow of the messages, determining which nodes are executed next. Edges can be *directed* or *conditional*. A *directed* edge represents a fixed, one-way flow of messages from one node to another. In contrast, a *conditional* edge introduces flexibility, allowing messages to reach different nodes based on a condition.

Fig. 1 illustrates the generalized representation of ChemGraph. First, the LLM agent is provided with a predefined set of tools (defined in the next section). Based on the user prompt, the agent decides which tool to invoke. After each tool call is finished, the LLM is provided with tool call result and decides if another tool call is needed. After all tool calls are completed, the message can follow one of the two paths. In the first path, ChemGraph acts like a conventional LLM, returns a human-like response that incorporates results from the tool calls. In the second mode, the messages are forwarded to a second LLM agent responsible for producing a structured output (with a predefined JSON schema) that directly addresses the user's request. This dual-mode design supports both natural language interaction and systematic evaluation, enabling ChemGraph to serve both typical use cases and evaluation workflows.

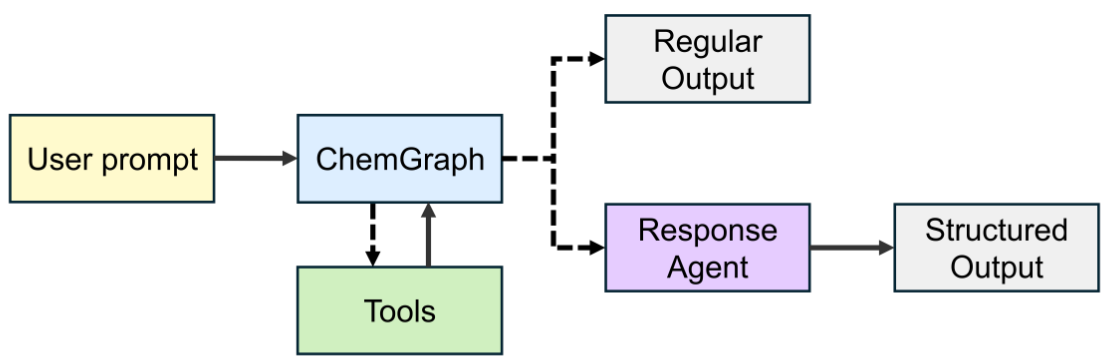

Figure 1: Overview of the ChemGraph. Solid arrows represent direct edges, indicating the default flow of messages. Dashed arrows represent conditional edges, where message flow depends on specific conditions. The figure lists the libraries used for workflow and cheminformatics, and for simulation backends.

**Tools.** In ChemGraph, tools are implemented as Python functions that wrap widely used libraries such as RDKit[33], Atomic Simulation Environment (ASE)[18], and PubChemPy[34] into structured and agent-compatible interfaces. The tools cover a broad range of functionalities, from basic cheminformatics operations such as converting a molecule names to a SMILES string and generating 3D molecular structure, to more advanced simulation workflows using ASE, including geometry optimization, vibrational frequency analysis, and thermochemistry calculations. Several tools use an internal data structure called AtomsData, which is a lightweight wrapper around the ASE Atoms class. It retains key information such as atomic numbers, atomic coordinates, cell parameters, and periodic boundary conditions. The primary purpose of AtomsData is to make atomic structure data serializable and compatible with LangGraph message passing. A summary of the tools utilized and evaluated in this study is presented in Table 1, with detailed specifications of their inputs and outputs available in Table S1 (supporting information).

Table 1: Tools integrated within ChemGraph, including their names and brief descriptions of their functionalities.

| Tool name | Tool Description |
| --- | --- |
| molecule_name_to_smiles | Convert a molecule name to a SMILES string |
| smiles_to_atomsdata | Convert a SMILES string to an AtomsData object |
| file_to_atomsdata | Convert a file to an AtomsData object |
| run_ase | Run molecular simulation (optimization, vibrational analysis, thermochemistry calculation) via ASE |
| save_atomsdata_to_file | Save an AtomsData object to a file (XYZ, mol, etc.) |
| calculator | Calculate a mathematical expression |

The use of ASE as our core simulation engine enables us to access a broad range of computational chemistry methods. Through its modular calculator interface, ChemGraph can access methodologies ranging from fast, low-cost approximations such as the extended tight-binding (xTB) method[35] and ML potentials like MACE[14,15], to more accurate but computationally intensive approaches, including DFT and post-Hartree-Fock methods via NWChem[19] and Orca[36,37]. This versatility allows users to balance computational cost and accuracy depending on the task, making the ChemGraph suitable for both high-throughput screening and high-accuracy electronic structure calculations.

**Prompt strategy.** Prompt plays an important role in establishing context and guiding the agent's behavior. In LangGraph's framework, the system prompt defines the available tools, the agent's responsibilities, and the expected structure of outputs. Kumar and colleagues have shown that having a modified the prompt to align the agent more with the domain of chemistry can help improve the success of tool calls.[26] In ChemGraph, we experimented with both general-purpose prompts and model-specific variants to optimize performance across different LLMs. These customized prompts included chemistry-specific language, general clarifications on tool input formats, and explicit instructions to avoid common failure modes (e.g., unnecessary tool calls or hallucinated tool outputs).

**Benchmarking and Evaluation.** We evaluated ChemGraph through 13 benchmark experiments involving six integrated tools, organized into three categories based on the type of user input: (1) molecule names, (2) SMILES strings and (3) chemical reactions. Table 2 summarizes the 260 independent evaluations conducted across the 13 experiments. In the first 11 experiments, which focus on tasks involving molecule names or SMILES, each requires up to four tool calls or subtasks. In contrast, the final two experiments, centered on computing thermochemical properties of chemical reactions, are significantly more complex, requiring between 9 and 12 tool calls depending on the number of reactants and products. These tasks involve executing thermochemical calculations for each species individually and create a final reaction-level property based on previous outputs.

Table 2: Summary of benchmark experiments used to evaluate ChemGraph. Each row includes the experiment label, a brief description, the number of subtasks, and the number of evaluation instances (e.g number of molecules, SMILES strings or reactions).

| Label | Description | Number of subtasks | Number of instances |
|---|---|---|---|
| name2smi | Convert a molecule name to a SMILES string | 1 | 30 |
| name2coord | Convert a molecule name to XYZ coordinates | 2 | 30 |
| name2opt | Run geometry optimization given a molecule name | 3 | 30 |
| name2vib | Run vibrational frequency given a molecule name | 3 | 15 |
| name2gibbs | Calculate the Gibbs free energy given molecule name at a temperature | 3 | 15 |
| names2file | Run geometry optimization and save the coordinates into a named XYZ file given a molecule name | 4 | 15 |
| smiles2coord | Convert a SMILES string to XYZ coordinates | 1 | 30 |
| smiles2opt | Run geometry optimization given a SMILES string | 2 | 30 |
| smiles2vib | Run vibrational frequency calculation given a SMILES string | 2 | 15 |
| smiles2gibbs | Calculate the Gibbs free energy of given SMILES string at a temperature | 2 | 15 |
| smiles2file | Run geometry optimization and save the coordinates into a named XYZ file given a SMILES string | 3 | 15 |
| react2enthalpy | Calculate reaction enthalpy at a temperature given a reaction | 9 to 12 | 10 |
| react2gibbs | Calculate the reaction Gibbs free energy at a temperature given a reaction | 9 to 12 | 10 |

For each experiment, we randomly selected a set of molecules from PubChem[34], setting a constraint on the maximum number of atoms per molecule (Tab. S2, supporting information). This constraint ensures each evaluation instance can be completed within a reasonable time, with stricter cutoffs for more computationally intensive tasks. The *name2opt* and *smiles2opt* tasks are exceptions; for these, we used a fixed list of small molecules across all runs because the geometry optimizations were performed using DFT.

We evaluated ChemGraph using a single-turn framework. Single-turn and multi-turn evaluations represent two different approaches to assessing performance of LLM agents.[38] In a single-turn evaluation, the agentic framework receives a user query and generates a complete response without human intervention. If the LLM makes a mistake in a tool call, it can attempt to self-correct within that single interaction. In contrast, multi-turn evaluation involves iterative human-AI interaction where humans provide corrective feedback when the AI makes mistakes, allowing for correction and refinement of the result through multiple exchanges. While ChemGraph allows multi-turn interaction and human-AI collaboration, multi-turn evaluation presents significant methodological challenges: different LLMs may exhibit different error patterns, leading to divergent correction paths and making it difficult to establish consistent and reproducible benchmarks.

To benchmark ChemGraph, we designed a standardized evaluation procedure with a template-based prompting approach. Each task was defined using a template prompt, with variable components, such as molecule names, SMILES strings or sets of reactants and products (Table S2, supporting information). These components are systematically varied in each instance of each experiment while keeping other components fixed, such as computation methods (e.g DFT, tight binding, ML potentials), conditions (e.g temperature), and file naming conventions. For every experiment and instance, we created a reference answer in structured JSON format. The reference answer represents how a domain expert would solve the problem based on the available tools, defining both the sequence of tool calls and expected results, which serve as ground truth for evaluation.

Performance of LLMs was evaluated based on two metrics: (1) the accuracy of the final answer, and (2) the number of tool calls used to reach that answer. To enable efficient and consistent benchmarking, we used ChemGraph's second operational mode, which generates structured outputs. We evaluated the performance of ChemGraph using four LLMs, categorized into open and proprietary models. The open model used was Qwen2.5-14B, a publicly released version of Alibaba's Qwen series. The proprietary models included GPT-4o-mini and GPT-4o from OpenAI, and Claude-3.5-haiku from Anthropic. We set the temperature of each mode to 0 to ensure consistency. However, we observe that even at zero temperature, the performance of the LLM still varies across different runs. Qwen2.5-14B was accessed through the Argonne Leadership Computing Facility (ALCF) using Globus Compute on ALCF's high-performance computing clusters. Access to the three proprietary models was provided via their respective APIs. Due to the high API cost associated with GPT-4o, its evaluation was limited to the two most complex tasks, *react2enthalpy* and *react2gibbs*.

**Multi-Agent System.** While a single LLM agent can handle simple tasks, its performance varies depending on the task complexity, especially when the workflow requires multiple, or chained tool calls. In contrast, a human expert can break down a complex workflow into smaller and more manageable subtasks. For instance, consider the simplified *react2enthalpy* task of calculating the reaction enthalpy for a reaction A+B→C. This problem can be broken down into three subtasks: compute the enthalpy of formation for A, B and C. Once those values are obtained, the overall reaction enthalpy can be determined.

Following this principle, we designed and evaluated a multi-agent version of ChemGraph for the two most complex tasks in this work, *react2enthalpy* and *react2gibbs*. Fig. 2 illustrates the multi-agent system's architecture. Instead of assigning the entire workflow to a single LLM, the multi-agent ChemGraph system comprises three main LLM agents: a Planner, an Executor, and an Aggregator agent. The Planner Agent decomposes the user's request into subtasks and generates a series of prompts, as demonstrated previously. These sub-prompts are passed to a Loop Controller, which sequentially forwards each prompt to an Executor agent. Each Executor is equipped with the same set of tools as in the single-agent system. After all subtasks are completed, their summarized results are sent to the Aggregator. The Aggregator integrates the original user query, the Planner's task decomposition, and summarized output from Executors to generate the final answer.

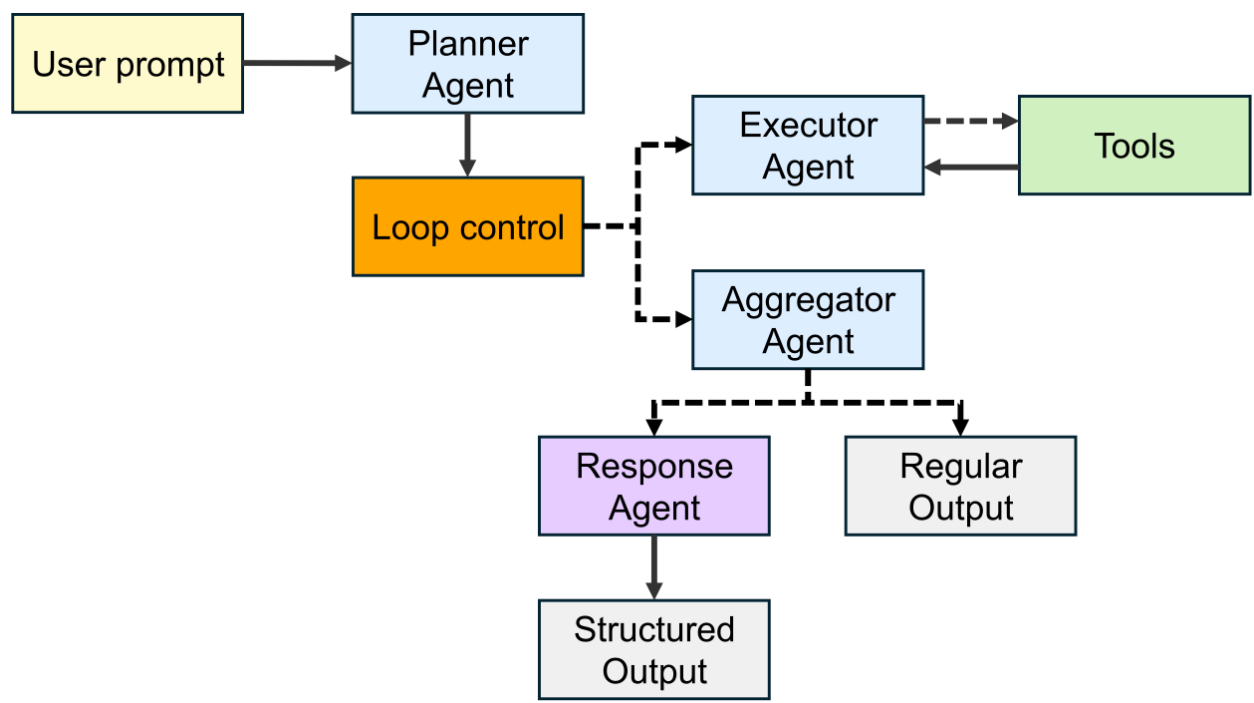

Figure 2: Architecture of the multi-agent ChemGraph. ChemGraph includes a series of specialized agents (Planner, Executor and Aggregator), coordinated by loop control logic. Outputs are generated in either regular or structured formats. Dashed arrows represent conditional edges, where message flow depends on specific conditions.

**Results**

**Single-agent evaluation.** We begin by evaluating the single-agent ChemGraph's performance on 13 experiments. Fig. 3 shows the accuracies of different models for different tasks. The accuracies of last two experiments, *react2enthalpy* and *react2gibbs*, are averaged over three independent runs. The task complexity, measured by the number of tool calls, is illustrated in Fig. 4.

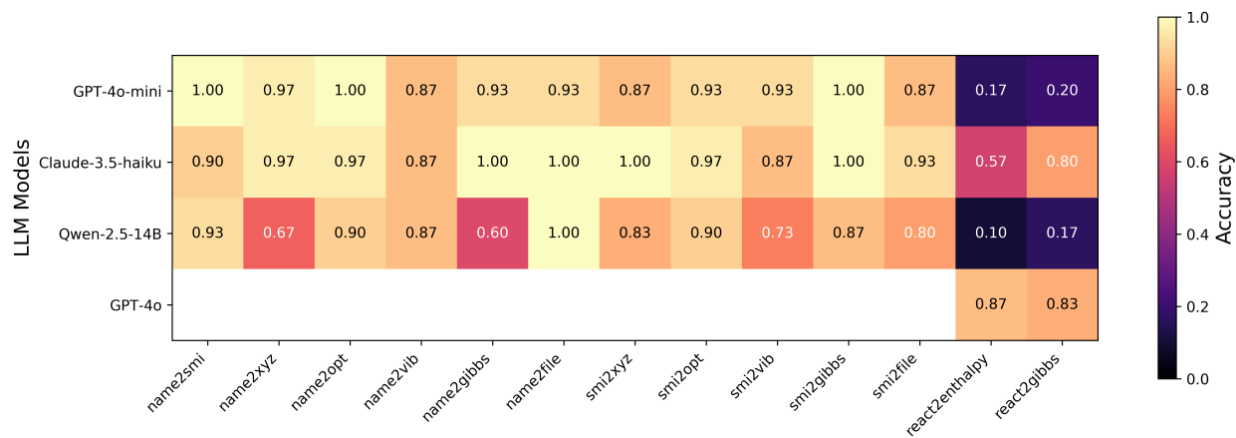

Figure 3: Accuracy of single-agent ChemGraph using different LLMs (GPT-4o-mini, Claude-3.5-haiku and Qwen2.5-14B) across 260 evaluation instances, grouped into 13 benchmark experiments. GPT-4o was only evaluated on 2 experiments, *react2enthalpy* and *react2gibbs*.

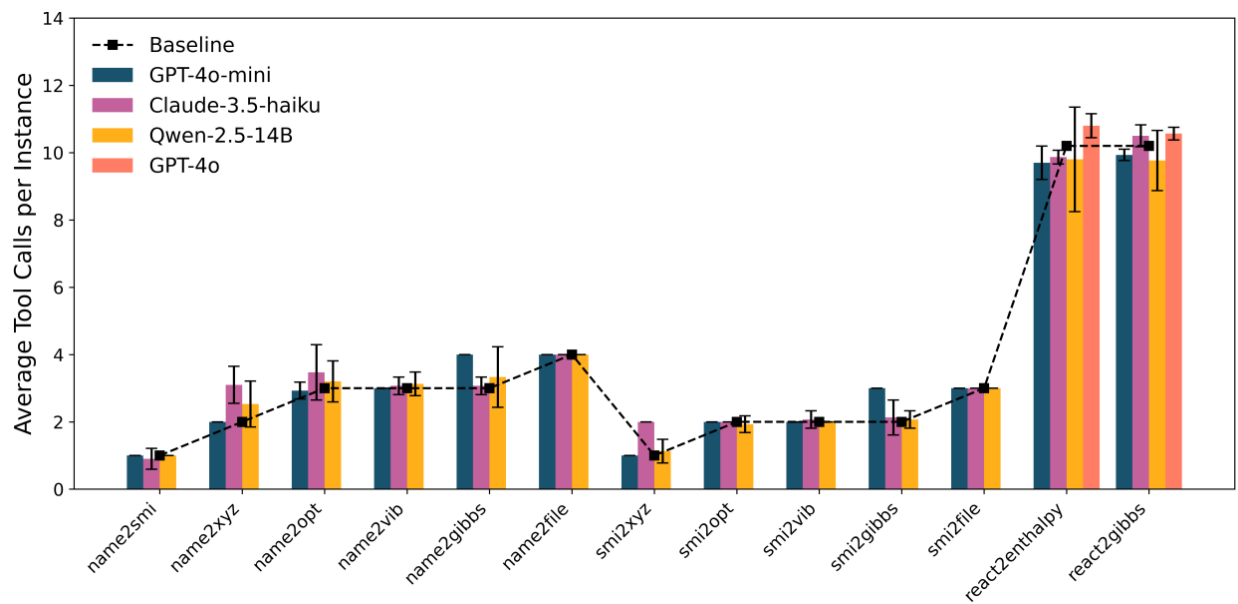

Figure 4: Number of tool calls made by different LLMs across 13 experiments, compared by baseline representing the number of tool calls required when solved manually by a human. GPT-4o was evaluated on 2 experiments, *react2enthalpy* and *react2gibbs*. Error bar representing the standard deviation of the number of tool calls per molecule for each experiment.

The first six experiments (*name2smi* to *name2file*) focus on cheminformatics and molecular simulation tasks using molecule names and other parameters as input. Both

Claude-3.5-haiku and GPT-4o-mini demonstrated strong and consistent performance, achieving over 87% accuracy in each task. Qwen2.5-14B showed more variability in performance, with an accuracy as low as 60% for the *name2gibbs* task.

For the *name2smi* task, Claude-3.5-haiku had the lowest performance among the models. Instead of consistently invoking the *molecule_name_to_smiles* tool, the model sometimes attempted to manually construct the SMILES strings using its internal chemical knowledge or declined to use the tool. Fig. 5 illustrates an example with a molecule 1-(2-methylphenyl)-3-(3-methylpyridin-2-yl)urea, where Claude-3.5-haiku declined to invoke *molecule_name_to_smiles* tool, assuming that the molecule's systematic name would not be compatible with it. Although the model can produce the correct output when explicitly prompted to use the tool, this requires additional user intervention. Because our evaluation focuses strictly on single-turn performance, the responses are still marked incorrect. The average number of tool calls per molecule for Claude-3.5-haiku for this task, as a result, is the lowest among the three models (Fig. 4). GPT-4o-mini accurately created correct answers for this experiment. Qwen2.5-14B generally produced accurate tool calls, including both tool names and arguments. However, its errors frequently occurred when extracting the results from the tool outputs.

**Human**

Provide the SMILES string corresponding to this molecule: 1-(2-methylphenyl)-3-(3-methylpyridin-2-yl)urea

**ChemGraph**

I'll help you find the SMILES string for the molecule 1-(2-methylphenyl)-3-(3-methylpyridin-2-yl)urea. To do this, I'll use the molecule_name_to_smiles function. However, in this case, the input is a systematic chemical name rather than a common molecule name, so the function might not work directly.

[…]

Would you like me to help you find alternative ways to represent this molecule?

**Human**

You should use molecule_name_to_smiles tool to obtain the SMILES string

**ChemGraph**

I apologize for my previous hesitation. You're right that I should attempt to use the molecule_name_to_smiles tool. However, the tool is designed for common molecule names, and the given name is a complex systematic name. Let me try to use the tool with the full systematic name

Tool Call:
    Tool name: molecule_name_to_smiles
    Args: 1-(2-methylphenyl)-3-(3-methylpyridin-2-yl)urea

Tool call output: "CC1=C(N=CC=C1)NC(=O)NC2=CC=CC=C2C"

Based on the tool output, the SMILES string for 1-(2-methylphenyl)-3-(3-methylpyridin-2-yl)urea is: CC1=C(N=CC=C1)NC(=O)NC2=CC=CC=C2C
[…]

Figure 5: Example multi-turn interaction with the ChemGraph agent in the *name2smi* task. The LLM initially failed to invoke the appropriate tool. After provided with human's feedback, it invoked the correct tools and reported the correct answer. For clarity, parts of the conversation have been omitted due to length, indicated by […].

The next three tasks, *name2xyz*, *name2opt*, and *name2vib*, require the LLMs to not only perform accurate tool calls but also generate correct and structured outputs. These outputs typically include a nested dictionary of atomic coordinates or a list containing both floating-point and complex numbers for vibrational frequency analysis. GPT-4o-mini and Claude-3.5-haiku performed well across these tasks. The most common errors we found across the models are errors in tool arguments, and issues during result summarization by the agent, such as missing values or confusion between float and complex number formats in the final report. Interestingly, for the *name2xyz* and *name2opt* tasks, Claude-3.5-haiku often invoke *save_atomsdata_to_file* tool after generating the coordinates, which led to a higher number of tool usage than the other two models (Figure 4 4).

For the *name2gibbs* task, GPT-4o-mini frequently omitted the calculator *parameter* when invoking the tool calls, causing initial failure. However, it frequently recovered using feedback from error messages achieving an accuracy of 93% (Fig. 3) but requiring more tool calls (average of 4).

The next five experiments (*smi2xyz* to *smi2file*) focus on tasks that use a SMILES string as input. Claude-3.5-haiku and GPT-4o-mini maintained strong and consistent performance, both achieving over 87% accuracy (Fig. 3). Qwen2.5-14B also showed improvement in this set, with a minimum accuracy of 73%. In general, we observe the same common errors and mistakes that also occurred in the first six experiments. These include incorrect tool call arguments and occasional errors when summarizing results, especially when dealing with complex outputs such as nested dictionaries or long lists of mixing floating-point and complex numbers.

In the last two tasks, *react2enthalpy* and *react2gibbs*, we also evaluated the performance for GPT-4o. To improve the reliability of the results, both accuracy (Fig. 3) and number of tool calls (Fig. 4) metrics for each model were averaged over three independent runs for each model. Both GPT-4o-mini and Qwen2.5-14B exhibited poor performance on these tasks, with accuracies not exceeding 20%. Claude-3.5-haiku gave a mixed performance (57% and 80% accuracy, respectively). GPT-4o performed the best, with an average accuracy above 83% for both tasks.

**Multi-agent evaluations.** In the previous experiments, LLMs generally performed well on tasks requiring up to four tool calls. However, their performance declined significantly in the final two experiments. As the number of tool calls and the amount of input and output increase, the LLM cannot retain relevant information, leading to incorrect tool calls, hallucinations, or data extraction failures. Reducing the context size for LLMs by including only essential information can mitigate this issue. For instance, when calculating the reaction enthalpy, the LLM only needs the enthalpies of formation for each species, not their full atomic coordinates or vibrational frequencies. To address these challenges, we implemented and evaluated a multi-agent version of ChemGraph (Fig. 5), where tasks were distributed among different agents (Fig. 2). In this setup, an executor agent handles all tool calls and summarizes intermediate results, while an aggregator agent uses only these summaries to compute final quantities, such as reaction enthalpy.

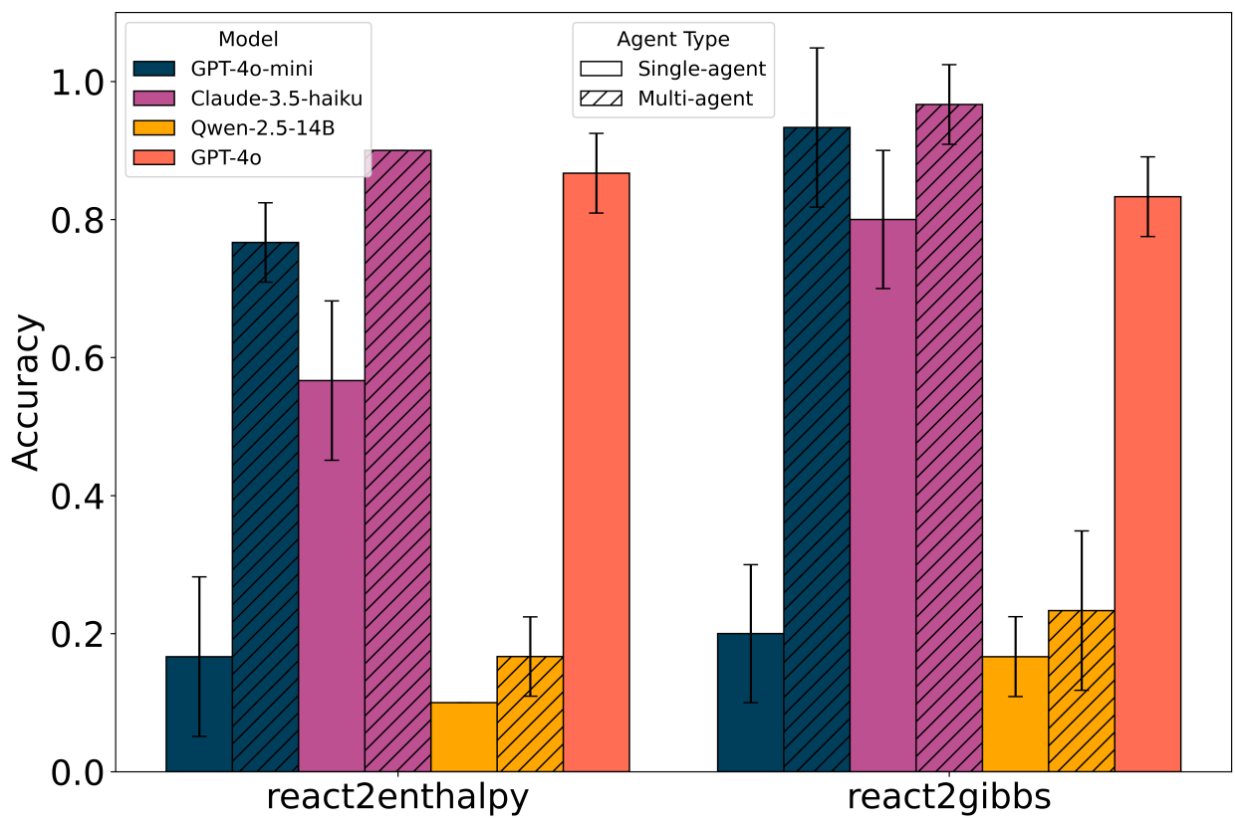

Figure 6: The average accuracy of multi-agent and single-agent ChemGraph using different LLMs for *react2enthalpy* and *react2gibbs* tasks. Error bars represent the standard deviation over three independent runs.

Fig. 6 demonstrates that the multi-agent system led to significant improvements. For the *react2enthalpy* task, GPT-4o-mini's accuracy increased from an average of 17% in the single-agent case to 77% across three independent runs. Claude-3.5-haiku showed a similar trend, improving from 57% to 90%, even surpassing GPT-4o's performance in single-agent case. Similar improvements were observed in the last task, *react2gibbs*, in which GPT-4o-mini performance improved from 20% to 93%, and Claude-3.5-haiku from 80% to 97%. In contrast, Qwen2.5-14B showed only modest gains, primarily due to persistent errors in individual tool calls, which limited the aggregator agent's ability to produce complete final answers. These results suggest that decomposing complex workflows into smaller, focused subtasks can substantially enhance LLM performance in molecular simulation workflows.

**Flexibility and Robustness of ChemGraph**. A key strength of our implementation lies in its adaptation of ASE calculators, which enables integration of a wide range of simulation packages and computational methods. Many modern simulation codes, including semi-empirical approaches like tight-binding, ML potentials such as MACE[14,15] and the Universal Models for Atoms (UMA)[16], to ab initio methods as implemented in NWChem and Orca, can be accessed through ASE calculators. This common layer allows ChemGraph to operate consistently across different simulation backends without changing its core logic. In our experiments, we demonstrated this flexibility by evaluating ChemGraph workflows using three different computational methods: DFT, tight-binding methods, and ML potentials. This modularity not only simplifies implementation but also enables rapid benchmarking and method substitution, depending on the user's desired trade-off between accuracy and computational cost.

We also successfully integrated UMA[16], a newly released ML atomic potential from Meta, with LangGraph. This showcases ChemGraph's ability to adapt quickly to emerging simulation technologies. As the ASE-compatible methods continue to grow, our framework is suited to incorporate new models and methods with minimal overhead.

**Limitations, safety and future directions.** ChemGraph is the initial release of our LLM agent system for computational chemistry and materials science workflows. As the LLM ecosystem advances, we plan to expand ChemGraph's capabilities by integrating new tools, enhancing agent design, and extending compatibility to a broader range of software platforms. A key goal of ChemGraph is to support execution of large-scale or long-running simulations. To that end, we are actively working on enabling automated generation of job submission scripts and deployment on high-performance computing (HPC) systems. This will empower users to efficiently manage large simulation campaigns across distributed environments.

Safety and reliability are also central to ChemGraph's design. LLM agents can sometimes behave unpredictably or make incorrect decisions. To mitigate this, we provide ChemGraph within a Docker container that ensures controlled execution environments, enforces safe dependencies, and limits unintended interactions with the host system or external resources.

While ChemGraph demonstrates strong performance in tool calling across our benchmark tasks, none of the four evaluated LLMs achieved perfect accuracy. Similar to human-operated workflows, LLMs still occasionally make errors. However, as LLM capabilities and their tool-calling capability continue to improve, we anticipate a corresponding boost in ChemGraph's effectiveness and robustness.

Our long-term vision is to make ChemGraph widely accessible to researchers, providing a natural language interface to high-quality simulation tools. One current limitation is the computational cost associated with using proprietary models. To address this, our results highlight the potential of combining ChemGraph with open-source models like Qwen2.5-14B, enabling cost-effective execution of complex scientific workflows. We believe the first version of ChemGraph provides a groundwork for systems of intelligent scientific agents that not only can automate simulations but also adapt, reason, and collaborate, paving the way for more efficient and accessible scientific discoveries.

**Conclusions**

In this work, we introduced ChemGraph, an LLM-powered agentic framework designed to automate molecular simulation workflows through structured tool calling and reasoning. We evaluated ChemGraph across a diverse set of 13 experiments, ranging from simple molecule-name-to-SMILES conversions to complex thermodynamic property calculations, and demonstrated that state-of-the-art LLMs can achieve high task completion accuracy.

We found that single-agent systems perform reliably on tasks involving a small number of tool calls. However, performance degrades for complex workflows due to context window saturation. To address this, we implemented a multi-agent version of ChemGraph that decomposes complex queries into smaller subtasks across agents. This approach significantly improved accuracy, especially for reaction enthalpy and Gibbs free energy calculations, where accuracy increased by up to 73% in small LLMs.

Another key strength of ChemGraph lies in its modular design and compatibility with ASE-style calculators, enabling integration of a wide range of simulation backends. We demonstrated ChemGraph's flexibility by executing workflows with DFT, tight binding and state-of-the-art machine learning potentials, highlighting its potential for rapid prototyping and benchmarking across computational methodologies. By coupling ChemGraph with fast and accurate machine learning potentials, we enable interactive and natural language-driven molecular simulations that support efficient and accurate exploration of chemical space.

**Data availability**

Evaluation data for ChemGraph is available on Zenodo:
https://zenodo.org/records/15579317.

**Code availability**

ChemGraph is available on GitHub: https://github.com/argonne-lcf/ChemGraph

**Acknowledgments**

This research used resources of the Argonne Leadership Computing Facility, a U.S. Department of Energy (DOE) Office of Science user facility at Argonne National Laboratory and is based on research supported by the U.S. DOE Office of Science-Advanced Scientific Computing Research Program, under Contract No. DE-AC02-06CH11357. Our work leverages ALCF Inference Endpoints, which provide a robust API for LLM inference on ALCF HPC clusters via Globus Compute. We are thankful to Serkan Altuntaş for his contributions to the user interface of ChemGraph and for insightful discussions on AIOps.